\documentclass[twocolumn]{openjournal}

\usepackage[english]{babel}

\makeatletter
\@ifundefined{languagealias}{%
  \@ifundefined{l@en}{\let\l@en\l@english}{}%
  \@ifundefined{captionsen}{}{}%
  \@ifundefined{dateen}{}{}%
  \@ifundefined{extrasen}{}{}%
  \@ifundefined{noextrasen}{}{}%
}{%
  \languagealias{en}{english}%
}
\makeatother
\usepackage{lipsum}

\usepackage{xcolor}
\usepackage{textgreek}
\usepackage[utf8]{inputenc}

\usepackage{enumitem}
\usepackage{svg}

\usepackage{hyperref}
\hypersetup{
    unicode, 
    colorlinks=true,
    linkcolor=linkcolor,
    citecolor=linkcolor,
    filecolor=linkcolor,
    urlcolor=linkcolor,
}
\usepackage{color,colortbl}
\definecolor{linkcolor}{rgb}{0.0,0.3,0.5}
\usepackage{tensind}
\tensordelimiter{?}
\DeclareGraphicsExtensions{.bmp,.png,.jpg,.pdf}
\usepackage{verbatim}

\usepackage{orcidlink}
\usepackage{soul}

\usepackage{graphicx}%
\usepackage{multirow}%
\usepackage{amsmath,amssymb,amsfonts}%
\usepackage{amsthm}%
\usepackage{mathrsfs}%
\usepackage[title]{appendix}%
\usepackage{xcolor}%
\usepackage{textcomp}%

\usepackage{booktabs}%
\usepackage{algorithm}%
\usepackage{algorithmicx}%
\usepackage{algpseudocode}%

\graphicspath{ {./figs/} }

\begin{document}

\title{HelioSpectrotron 5000 \\ An Interactive Multi-Resolution Solar Spectral Atlas }

\author{\vspace*{-35px}Alexander G.M. Pietrow\orcidlink{0000-0002-0484-7634}}
\email{apietrow@aip.de}
\affiliation{Leibniz-Institut für Astrophysik Potsdam (AIP), An der Sternwarte 16, 14482 Potsdam, Germany}

\begin{abstract}
\href{http://hs5000.vo.aip.de/}{HelioSpectrotron~5000} (HS5000) is an interactive, multi-resolution solar spectral atlas designed to facilitate direct comparison between high-resolution reference spectra and observations obtained with a wide range of ground-based instruments. Based on the Hamburg FTS atlas, the HS5000 provides both absolute intensity and continuum-normalized spectra at arbitrary spectral ranges and resolutions, as well as curated line identifications and optional telluric contamination. This framework enables rapid wavelength calibration, line identification, and context image generation.\vspace*{10px}
\end{abstract}


\maketitle

\section{Introduction}
\label{sec:intro}

Solar reference spectra, also known as solar atlases, provide high-resolution, high signal-to-noise representations of the solar spectrum.  Constructed either from averaged disk-center observations (flux atlases) or from observations integrated over the full solar disk, known as full-disk or Sun-as-a-star atlases (See \citet{Doerr2016} and \citet{Foad2025} for a detailed review). Initially, such atlases have served as essential tools for identifying and navigating the highly structured solar spectrum \citep[][]{Fraunhofer1817}, however, over time their function has shifted well beyond this use. Modern solar atlases are now widely employed for, among other things, wavelength and intensity calibration \citep[e.g.][]{Lofdahl2021A&A...653A..68L}, and as reference benchmarks for studies of radial-velocity variability \citep{Trifonov20, AlMoulla2024, Lakeland2024}, convective signatures \citep[e.g.][]{Meunier2010A&A...519A..66M, 2023A&A...680A..62E, Pietrow23}, and the magnetic activity \citep[e.g.][]{Cretignier24, Pietrow2023Nessi, DeWilde2025, Foad2025}.

\begin{figure*}
    \centering
    \includegraphics[width=1\linewidth]{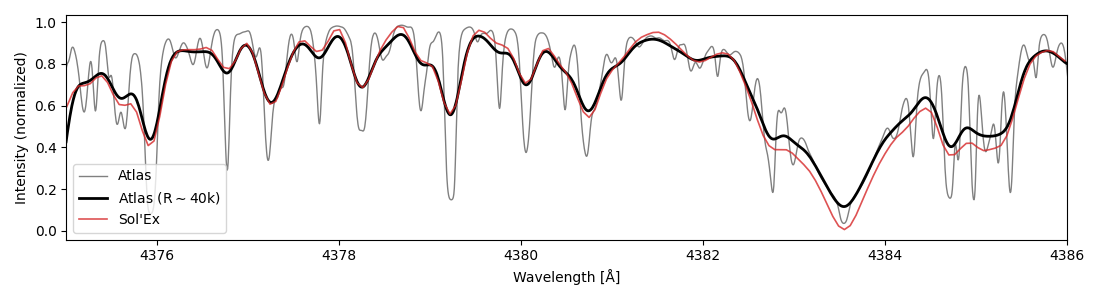}
    \caption{Example Sol’Ex spectrum (red) taken from \citet{VaradiNagy2025} compared to the FTS atlas (gray) and a version of the same atlas convolved to the spectral resolution of the Sol’Ex (black), illustrating how spectral alignment becomes non-trivial when the spectral resolutions differ strongly.}
    \label{fig:resolution}
    \vspace{0.5cm}
\end{figure*}

Despite the availability of numerous high-quality solar datasets and analysis tools addressing many of these applications, solar atlases as general spectral reference frameworks remain largely confined to static books and tabulated resources. At the same time, recent developments have renewed the need for accessible and navigable reference spectra at specific spectral resolutions (See Fig.~\ref{fig:resolution}). In particular, the emergence of dedicated Sun-as-a-star facilities delivering synoptic, disk-integrated spectra over broad wavelength ranges \citep[e.g.][]{2023Zhao}, the forthcoming Paranal Solar ESPRESSO Telescope \citep[POET;][]{Santos2025}, and the rapid uptake of low-cost amateur spectrographs such as the Solar Explorer \citep[Sol'Ex;][]{2023Buil} have opened new observational parameter spaces that previously required specialized instrumentation or narrowband filters to access.

Currently, the only interactive reference atlas is provided by the Base de Données Solaire Sol \citep[BASS2000, ][]{Aboudarham2000}\footnote{\url{https://bass2000.obspm.fr/solar_spect.php}}. This web-based application spanning wavelengths from approximately 670 to 540,000~Å is an amalgamation of the ultraviolet atlas of \citet{Curdt2001}, the visible Liège atlas \citep{Delbouille1973}, and the infrared Liège atlas \citep{Delbouille1963}. Additionally, spectral line identification is provided via integration with the  NIST Atomic Spectra Database \citep{Kramida1991}. However, this approach frequently returns incorrect identifications, as the tabulated line strengths in the database do not correlate reliably with observed line depths in solar spectra. As a result, prominent features can be misidentified\footnote{ For example, the H$\delta$ line is erroneously assigned to an iron transition.}.
So, while this resource represents an important step toward interactive atlas usage, it lacks several features required for comprehensive application to modern solar spectroscopy for both professionals and amateurs alike. Several other digital datasets are made publicly available, such as the two atlases by the Institut für Astrophysik und Geophysik\footnote{\url{https://www.astro.physik.uni-goettingen.de/research/flux_atlas/}}$^,$\footnote{\url{https://www.astro.physik.uni-goettingen.de/research/solar-lib/}}  \citep{Reiners2016, Ellwarth2023}, however these primarily represent static digital renditions of their underlying datasets, rather than fully interactive spectral reference tools.

To function effectively as a calibration and line-identification resource for solar instrumentation, a reference atlas should satisfy the following requirements: 

\begin{enumerate}[label=\roman*,align=left]
\item Cover the full wavelength range accessible from the ground.

\item Be available both as continuum-normalized spectra and in absolute intensity units.

\item Include reliable, well-curated spectral line identifications.

\item Allow flexible extraction of user-defined wavelength intervals.

\item Provide both one-dimensional spectra and two-dimensional simulated slit representations, facilitating navigation and interpretation of raw two-dimensional spectral data.

\item Be available at multiple spectral resolutions, ranging from Fourier-transform spectrometer quality down to the lower resolving powers characteristic of instruments such as Sol’Ex.

\item Account for varying levels of telluric absorption, enabling realistic comparison with observations obtained under different atmospheric conditions and observing sites.
\end{enumerate}

In this work, I present the HelioSpectrotron 5000 (HS5000), a tool designed explicitly to meet these requirements. 

\section{Methods and implementation}\label{sec:methods}
Over the last decades, a wide range of solar spectral atlases has been produced, spanning different wavelength ranges, spectral resolutions, calibration strategies, and activity levels (see \citet{Foad2025} for a detailed comparison). Because our analysis requires access to both continuum-normalized spectra and absolute fluxes, the most suitable reference is one of the Kitt Peak Fourier Transform Spectrometer (FTS) atlases known as the Hamburg atlas \citep[][]{Neckel1984}, as it provides a high signal-to-noise ratio together with consistently calibrated absolute and normalized flux scales over a wide range.

This atlas, covering 3290 -- 12,500~\AA, has a spectral resolution of R $ \sim$ 400,000. It is an averaged spectrum taken over six days between November 1980 and June 1981 with the 1.6m McMath–Pierce Solar Telescope \citep{Pierce1964}, and is one of the most widely used solar atlases to date. This work uses the version built into the `atlas' subpackage of the \textit{ISPy} Python library \citep{ISPy2021}. 

A set of two telluric spectra including all atmospheric elements is obtained from the \textit{Transmissions Atmosphériques Personnalisées pour l'AStronomie} \citep[TAPAS, ][]{Bertaux2014} code\footnote{\url{https://tapas.aeris-data.fr/en/home/}}. In both cases, the data were taken at a zenith angle of 45$^\circ$ (airmass 1.412) using the default TAPAS atmosphere and mixing ratios. The only difference was the observatory height, which was set to sea level (0~m) and 2500~m, respectively. These spectra are generated between 300 and 12,000~\AA, thus clipping the red end of the atlas. 

Both the atlas and one of the telluric spectra are plotted for a given window (25~\AA\ by default), and the product of the two is transformed into a simulated slit spectrum by tiling the resulting spectrum 120 times. These 1D and 2D spectra can be degraded to lower resolutions by convolving with a Gaussian kernel. In the normalized representation, the continuum level may appear slightly below unity, since the telluric transmission is applied multiplicatively and includes wavelength-dependent continuum absorption. However this multiplication can be turned off when intensities are required.

Wavelengths are given in air by default, but can be converted to approximate vacuum wavelengths\footnote{The refractive index of air depends on atmospheric conditions and altitude and therefore does not permit an exact conversion.} using the prescription of \citet{Greisen2006}, as implemented in \textit{specutils} \citep{specutils2019}.

For line identification, the classical line lists by \citet{Moore1996} and \citet{Babcock1947} are adopted. These compilations cover the wavelength ranges 2935–8770~\AA\ and 6600–13495~\AA, respectively, and were explicitly constructed to identify the deepest lines in the solar spectrum. As these tables are available only in printed form, they were digitized using \textit{extracttable}\footnote{\url{https://www.extracttable.com/}} and post-processed to obtain a machine-readable line list where only the first element was kept if multiple elements were suggested for that wavelength. 
As the labels serve mostly to aid the user in calibrating their instrument, a sparser but more accurate list is preferred over a denser list with uncertainties. For this reason, all ambiguous entries were removed after OCR related-errors in ionization and element notation were corrected.
Afterwards, the deepest line for each 0.5~\AA\ interval was selected, and the remaining set of about 6000 lines was painstakingly cross-matched by hand against the original source. These lists and the source code have been made available in the project \href{https://github.com/AlexPietrow/HelioSpectrotron5000}{Github}\footnote{\url{https://github.com/AlexPietrow/HelioSpectrotron5000}}. 

\begin{figure*}
    \centering
    \includegraphics[width=1\linewidth]{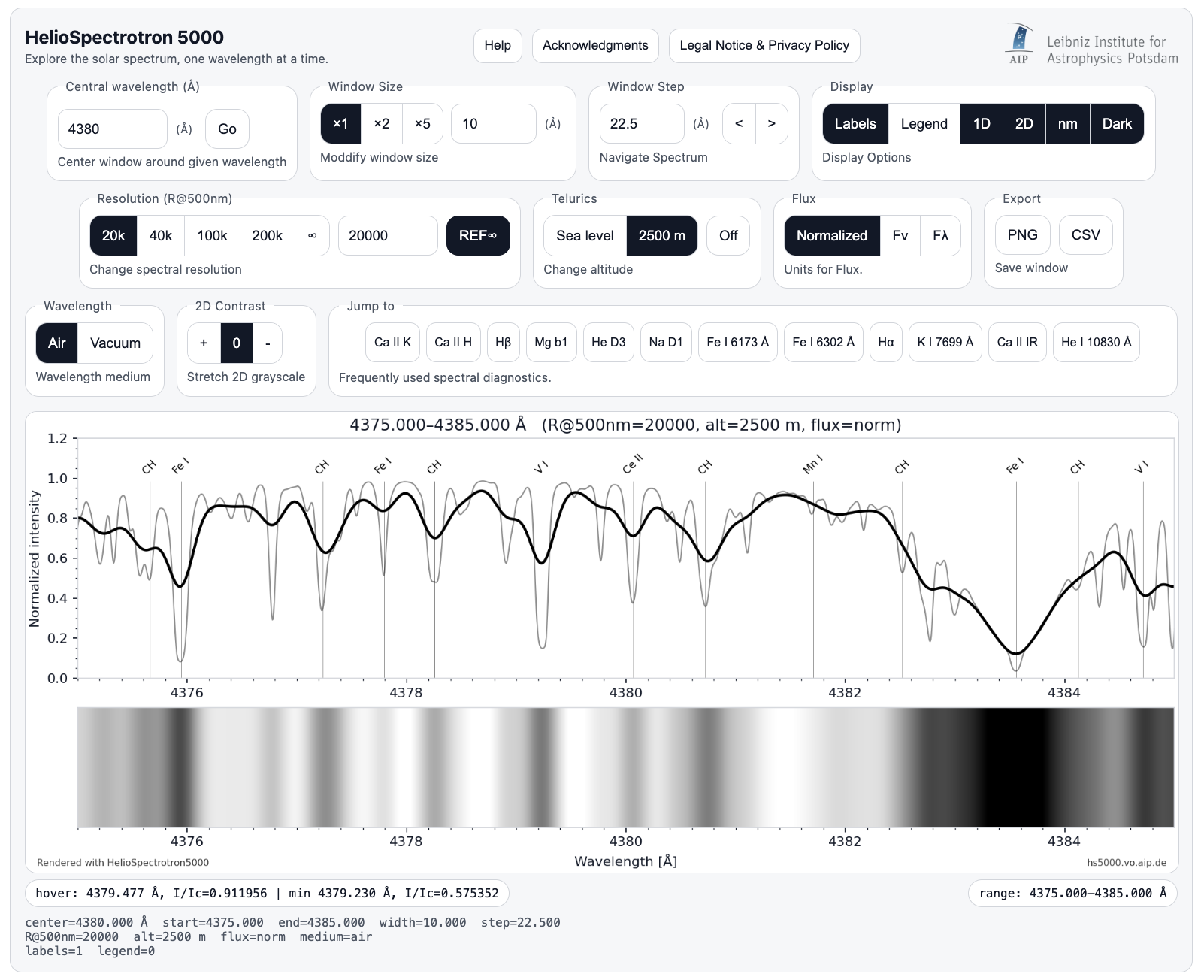}
    \caption{Overview of HS5000 showing the full user interface, tuned to the same spectral range as Fig.~\ref{fig:resolution}. The native-resolution atlas spectrum (gray) is compared to a degraded spectrum at $R \approx 20{,}000$ (black). Line labels are enabled, and the associated 2D spectrum panel is shown for navigation. Tellurics are turned off.\vspace{10px}}
    \label{fig:overview}
    \vspace{0.5cm}
\end{figure*}

\begin{figure*}
    \centering
    \includegraphics[width=1\linewidth]{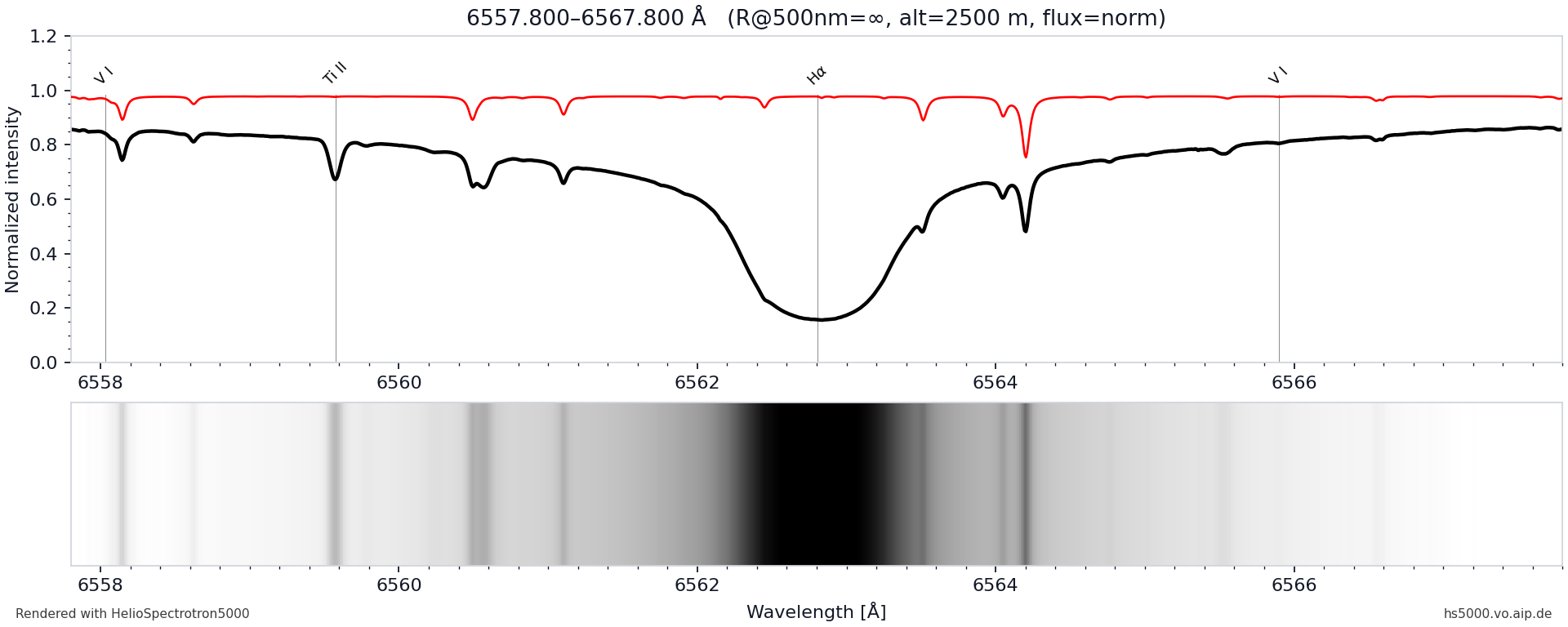}
    \caption{A full resolution spectrum around the H$\alpha$ line with tellurics set to approximate observations at sea level.}
    \label{fig:halpha}
\end{figure*}

The final product is rendered through a lightweight web application consisting of a Python-based $uvicorn$ backend built around `FastAPI'\footnote{\url{https://fastapi.tiangolo.com/}} and an HTML/JavaScript frontend, which allows for an interactive atlas with minimal server-side processing\footnote{The HS5000 can be found at \url{hs5000.vo.aip.de}, with a mirror at \url{atlas.pietrow.net}.}.

With this implementation of the HS5000 atlas, calibrations such as the one shown in Fig.~\ref{fig:resolution} become straightforward, as the user interface (UI) enables the construction of spectra at arbitrary spectral resolution. The resulting spectra can be exported directly as numerical arrays and compared quantitatively with observed data. Rough wavelength values can be obtained by moving the mouse\footnote{The backend communicates the wavelength mapping to the frontend by embedding a narrow encoded strip at the bottom of each generated image.}, and the raw data used to create the plot can be downloaded for further processing. 

An overview of the full UI is presented in Fig.~\ref{fig:overview}, illustrating a low-resolution spectrum representative of instruments such as Sol’Ex. In this configuration, the accompanying 2D spectrum facilitates rapid identification of the corresponding wavelength range by approximating a slitjaw context image.

A contrasting use case is shown in Fig.~\ref{fig:halpha}, where strong telluric absorption corresponding to sea-level conditions is applied multiplicatively to the H$\alpha$ spectral region. This example demonstrates the impact of atmospheric transmission on the line shape.

\section{Conclusions}\label{sec:Conclusion}

This work presents the \textit{HelioSpectrotron~5000} (HS5000), an interactive, multi-resolution solar atlas designed to serve as a flexible reference framework for modern solar spectroscopy. The primary strength of this atlas lies in its ability to brige the gap between static reference atlases and real observational data. Built on the high-resolution \citep{Neckel1984} FTS atlas, this application provides arbitrary spectral resolutions, telluric transmission, and curated line identification. This allows the HS5000 to function as a direct and robust instrument calibration and spectral identification tool. The option to export spectra as arrays further enables quantitative calibrations and custom plots.

The HS5000 is intended both for professional applications, such as wavelength calibration and instrument characterization, and for advanced amateur spectroscopy, where access to resolution-matched reference spectra has historically been limited. While the present implementation focuses on disk-center spectra and ground-based wavelength ranges, the framework is designed to be extensible.

Future developments will include disk-integrated representations suitable for solar--stellar comparisons, and potentially an extended wavelength range. In its current form, the HS5000 is a practical and accessible tool that complements existing solar atlases by emphasizing interactivity, reproducibility, and ease of use.

\section{Website}
The atlas can be found at \url{hs5000.vo.aip.de}, with a mirror at \url{atlas.pietrow.net}.

\section*{Acknowledgments}
AP is supported by the \emph{Deut\-sche For\-schungs\-ge\-mein\-schaft, DFG\/} project number PI 2102/1-1
AP acknowledges productive discussions with, and beta testing by, Pál Váradi Nagy and Carsten Denker. Additionally, AP thanks Harry Enke for his assistance with local website hosting, Rebecca Klee for her help with copyright questions, as well as Casper Pietrow for providing a conducive HOMM3 
environment during the preparation of this manuscript. Additionally, AP thanks the anonymous referee for their valuable suggestions during the peer-review process.
This research has made use of NASA's Astrophysics Data System (ADS) bibliographic services. 

\bibliographystyle{apsrev4-1}

\bibliography{main}

@ARTICLE{Foad2025,
       author = {{Hanassi-Savari}, F. and {Pietrow}, A.~G.~M. and {Druett}, M.~K. and {Cretignier}, M. and {Ellwarth}, M.},
        title = "{Solar flux atlases: The new HARPS-N quiet Sun benchmark and continuum normalisation of the Ca II H \& K lines}",
      journal = {\aap},
         year = 2025,
        month = oct,
       volume = {702},
          eid = {A97},
        pages = {A97},
          doi = {10.1051/0004-6361/202451874},
archivePrefix = {arXiv},
       eprint = {2508.07912},
 primaryClass = {astro-ph.SR},
       adsurl = {https://ui.adsabs.harvard.edu/abs/2025A&A...702A..97H}
}

@ARTICLE{2023Buil,
       author = {{Buil}, Christian and {Malherbe}, Jean-Marie and {Maksimovic}, Milan},
        title = "{Sol'Ex et l'imagerie monochromatique solaire}",
      journal = {Photoniques},
         year = 2023,
        month = jul,
       volume = {120},
        pages = {36-40},
          doi = {10.1051/photon/202312036},
       adsurl = {https://ui.adsabs.harvard.edu/abs/2023Phot..120...36B}
}

@BOOK{Delbouille1973,
       author = {{Delbouille}, L. and {Roland}, G. and {Neven}, L.},
        title = "{Atlas photometrique du spectre solaire de [lambda] 3000 a [lambda] 10000}",
         year = 1973,
       adsurl = {https://ui.adsabs.harvard.edu/abs/1973apds.book.....D}
}

@ARTICLE{2023Zhao,
       author = {{Zhao}, Lily L. and {Dumusque}, Xavier and {Ford}, Eric B. and {Llama}, Joe and {Mortier}, Annelies and {Bedell}, Megan and {Al Moulla}, Khaled and {Bender}, Chad F. and {Blake}, Cullen H. and {Brewer}, John M. and {Collier Cameron}, Andrew and {Cosentino}, Rosario and {Figueira}, Pedro and {Fischer}, Debra A. and {Ghedina}, Adriano and {Gonzalez}, Manuel and {Halverson}, Samuel and {Kanodia}, Shubham and {Latham}, David W. and {Lin}, Andrea S.~J. and {Lo Curto}, Gaspare and {Lodi}, Marcello and {Logsdon}, Sarah E. and {Lovis}, Christophe and {Mahadevan}, Suvrath and {Monson}, Andrew and {Ninan}, Joe P. and {Pepe}, Francesco and {Roettenbacher}, Rachael M. and {Roy}, Arpita and {Santos}, Nuno C. and {Schwab}, Christian and {Stef{\'a}nsson}, Gu{\dj}mundur and {Szymkowiak}, Andrew E. and {Terrien}, Ryan C. and {Udry}, Stephane and {Weiss}, Sam A. and {Wildi}, Fran{\c{c}}ois and {Wildi}, Thibault and {Wright}, Jason T.},
        title = "{The Extreme Stellar-signals Project. III. Combining Solar Data from HARPS, HARPS-N, EXPRES, and NEID}",
      journal = {\aj},
         year = 2023,
        month = oct,
       volume = {166},
       number = {4},
          eid = {173},
        pages = {173},
          doi = {10.3847/1538-3881/acf83e},
archivePrefix = {arXiv},
       eprint = {2309.03762},
 primaryClass = {astro-ph.EP},
       adsurl = {https://ui.adsabs.harvard.edu/abs/2023AJ....166..173Z}
}

@ARTICLE{Ellwarth2023,
       author = {{Ellwarth}, M. and {Sch{\"a}fer}, S. and {Reiners}, A. and {Zechmeister}, M.},
        title = "{The IAG spectral atlas of the spatially resolved Sun: Centre-to-limb observations}",
      journal = {\aap},
         year = 2023,
        month = may,
       volume = {673},
          eid = {A19},
        pages = {A19},
          doi = {10.1051/0004-6361/202245612},
archivePrefix = {arXiv},
       eprint = {2303.08205},
 primaryClass = {astro-ph.SR},
       adsurl = {https://ui.adsabs.harvard.edu/abs/2023A&A...673A..19E}
}

@ARTICLE{Doerr2016,
       author = {{Doerr}, H. -P. and {Vitas}, N. and {Fabbian}, D.},
        title = "{How different are the Li{\`e}ge and Hamburg atlases of the solar spectrum?}",
      journal = {\aap},
         year = 2016,
        month = may,
       volume = {590},
          eid = {A118},
        pages = {A118},
          doi = {10.1051/0004-6361/201628570},
archivePrefix = {arXiv},
       eprint = {1604.03748},
 primaryClass = {astro-ph.SR},
       adsurl = {https://ui.adsabs.harvard.edu/abs/2016A&A...590A.118D}
}

@ARTICLE{Lofdahl2021A&A...653A..68L,
       author = {{L{\"o}fdahl}, Mats G. and {Hillberg}, Tomas and {de la Cruz Rodr{\'\i}guez}, Jaime and {Vissers}, Gregal and {Andriienko}, Oleksii and {Scharmer}, G{\"o}ran B. and {Haugan}, Stein V.~H. and {Fredvik}, Terje},
        title = "{SSTRED: Data- and metadata-processing pipeline for CHROMIS and CRISP}",
      journal = {\aap},
         year = 2021,
        month = sep,
       volume = {653},
          eid = {A68},
        pages = {A68},
          doi = {10.1051/0004-6361/202141326},
archivePrefix = {arXiv},
       eprint = {1804.03030},
 primaryClass = {astro-ph.IM},
       adsurl = {https://ui.adsabs.harvard.edu/abs/2021A&A...653A..68L}
}

@article{VaradiNagy2025,
  author       = {P. {Váradi Nagy} and A. G. M. Pietrow},
  title        = {An Atlas of Spectroheliograms from 3641 to 6600 Å},
  journal      = {Research Notes of the AAS},
  volume       = {9},
  number       = {7},
  pages        = {188},
  year         = {2025},
  doi          = {10.3847/2515-5172/adef50},
  url          = {https://doi.org/10.3847/2515-5172/adef50}
}

@ARTICLE{AlMoulla2024,
       author = {{Al Moulla}, K. and {Dumusque}, X. and {Cretignier}, M.},
        title = "{Measuring precise radial velocities on individual spectral lines. IV. Stellar activity correlation with line formation temperature}",
      journal = {\aap},
         year = 2024,
        month = mar,
       volume = {683},
          eid = {A106},
        pages = {A106},
          doi = {10.1051/0004-6361/202348150},
archivePrefix = {arXiv},
       eprint = {2312.10836},
 primaryClass = {astro-ph.SR},
       adsurl = {https://ui.adsabs.harvard.edu/abs/2024A&A...683A.106A}
}

@ARTICLE{Lakeland2024,
       author = {{Lakeland}, Ben S. and {Naylor}, Tim and {Haywood}, Rapha{\"e}lle D. and {Meunier}, Nad{\`e}ge and {Rescigno}, Federica and {Dalal}, Shweta and {Mortier}, Annelies and {Thompson}, Samantha J. and {Cameron}, Andrew Collier and {Dumusque}, Xavier and {L{\'o}pez-Morales}, Mercedes and {Pepe}, Francesco and {Rice}, Ken and {Sozzetti}, Alessandro and {Udry}, St{\'e}phane and {Ford}, Eric and {Ghedina}, Adriano and {Lodi}, Marcello},
        title = "{The magnetically quiet solar surface dominates HARPS-N solar RVs during low activity}",
      journal = {\mnras},
         year = 2024,
        month = jan,
       volume = {527},
       number = {3},
        pages = {7681-7691},
          doi = {10.1093/mnras/stad3723},
archivePrefix = {arXiv},
       eprint = {2311.16076},
 primaryClass = {astro-ph.SR},
       adsurl = {https://ui.adsabs.harvard.edu/abs/2024MNRAS.527.7681L}
}

@ARTICLE{Reiners2016,
       author = {{Reiners}, A. and {Mrotzek}, N. and {Lemke}, U. and {Hinrichs}, J. and {Reinsch}, K.},
        title = "{The IAG solar flux atlas: Accurate wavelengths and absolute convective blueshift in standard solar spectra}",
      journal = {\aap},
         year = 2016,
        month = mar,
       volume = {587},
          eid = {A65},
        pages = {A65},
          doi = {10.1051/0004-6361/201527530},
archivePrefix = {arXiv},
       eprint = {1511.03014},
 primaryClass = {astro-ph.SR},
       adsurl = {https://ui.adsabs.harvard.edu/abs/2016A&A...587A..65R}
}

@INPROCEEDINGS{ISPy2021,
       author = {{Diaz Baso}, C. and {Vissers}, G. and {Calvo}, F. and {Pietrow}, A.~G.~M. and {Yadav}, R. and {de la Cruz Rodr{\'\i}guez}, J. and {Zivadinovic}, L.},
        title = "{ISPy}",
    booktitle = {Zenodo Software package},
         year = 2021,
       volume = {56},
        month = oct,
          eid = {5608441},
        pages = {5608441},
          doi = {10.5281/zenodo.5608441},
       adsurl = {https://ui.adsabs.harvard.edu/abs/2021zndo...5608441D}
}

@ARTICLE{Trifonov20,
       author = {{Trifonov}, Trifon and {Tal-Or}, Lev and {Zechmeister}, Mathias and {Kaminski}, Adrian and {Zucker}, Shay and {Mazeh}, Tsevi},
        title = "{Public HARPS radial velocity database corrected for systematic errors}",
      journal = {\aap},
         year = 2020,
        month = apr,
       volume = {636},
          eid = {A74},
        pages = {A74},
          doi = {10.1051/0004-6361/201936686},
archivePrefix = {arXiv},
       eprint = {2001.05942},
 primaryClass = {astro-ph.EP},
       adsurl = {https://ui.adsabs.harvard.edu/abs/2020A&A...636A..74T}
}

@ARTICLE{Pietrow23,
       author = {{Pietrow}, A.~G.~M. and {Kiselman}, D. and {Andriienko}, O. and {Petit dit de la Roche}, D.~J.~M. and {D{\'\i}az Baso}, C.~J. and {Calvo}, F.},
        title = "{Center-to-limb variation of spectral lines and continua observed with SST/CRISP and SST/CHROMIS}",
      journal = {\aap},
         year = 2023,
        month = mar,
       volume = {671},
          eid = {A130},
        pages = {A130},
          doi = {10.1051/0004-6361/202244811},
archivePrefix = {arXiv},
       eprint = {2212.03991},
 primaryClass = {astro-ph.SR},
       adsurl = {https://ui.adsabs.harvard.edu/abs/2023A&A...671A.130P}
}

@ARTICLE{Meunier2010A&A...519A..66M,
       author = {{Meunier}, N. and {Lagrange}, A. -M. and {Desort}, M.},
        title = "{Reconstructing the solar integrated radial velocity using MDI/SOHO}",
      journal = {\aap},
         year = 2010,
        month = sep,
       volume = {519},
          eid = {A66},
        pages = {A66},
          doi = {10.1051/0004-6361/201014199},
archivePrefix = {arXiv},
       eprint = {1005.4764},
 primaryClass = {astro-ph.SR},
       adsurl = {https://ui.adsabs.harvard.edu/abs/2010A&A...519A..66M}
}

@ARTICLE{Cretignier24,
       author = {{Cretignier}, M. and {Pietrow}, A.~G.~M. and {Aigrain}, S.},
        title = "{Stellar surface information from the Ca II H\&K lines - I. Intensity profiles of the solar activity components}",
      journal = {\mnras},
         year = 2024,
        month = jan,
       volume = {527},
       number = {2},
        pages = {2940-2962},
          doi = {10.1093/mnras/stad3292},
archivePrefix = {arXiv},
       eprint = {2310.15926},
 primaryClass = {astro-ph.SR},
       adsurl = {https://ui.adsabs.harvard.edu/abs/2024MNRAS.527.2940C}
}

@INPROCEEDINGS{Pietrow2023Nessi,
       author = {{Pietrow}, Alexander G.~M. and {Pastor Yabar}, Adur},
        title = "{Center-to-limb variation of spectral lines and their effect on full-disk observations}",
    booktitle = {Dynamics of Solar and Stellar Convection Zones and Atmospheres},
         year = 2024,
       editor = {{Getling}, Alexander V. and {Kitchatinov}, Leonid L.},
       series = {IAU Symposium},
       volume = {365},
        month = dec,
        pages = {389-393},
          doi = {10.1017/S174392132300501X},
archivePrefix = {arXiv},
       eprint = {2311.06200},
 primaryClass = {astro-ph.SR},
       adsurl = {https://ui.adsabs.harvard.edu/abs/2024IAUS..365..389P}
}

@ARTICLE{2023A&A...680A..62E,
       author = {{Ellwarth}, M. and {Ehmann}, B. and {Sch{\"a}fer}, S. and {Reiners}, A.},
        title = "{Convective characteristics of Fe I lines across the solar disc}",
      journal = {\aap},
         year = 2023,
        month = dec,
       volume = {680},
          eid = {A62},
        pages = {A62},
          doi = {10.1051/0004-6361/202347615},
archivePrefix = {arXiv},
       eprint = {2310.15782},
 primaryClass = {astro-ph.SR},
       adsurl = {https://ui.adsabs.harvard.edu/abs/2023A&A...680A..62E}
}

@ARTICLE{Curdt2001,
       author = {{Curdt}, W. and {Brekke}, P. and {Feldman}, U. and {Wilhelm}, K. and {Dwivedi}, B.~N. and {Sch{\"u}hle}, U. and {Lemaire}, P.},
        title = "{The SUMER spectral atlas of solar-disk features}",
      journal = {\aap},
         year = 2001,
        month = aug,
       volume = {375},
        pages = {591-613},
          doi = {10.1051/0004-6361:20010364},
       adsurl = {https://ui.adsabs.harvard.edu/abs/2001A&A...375..591C}
}

@BOOK{Delbouille1963,
       author = {{Delbouille}, Luc and {Roland}, Ginette},
        title = "{Atlas photometrique du spectre solaire de [lambda] 7498 a [lambda] 12016. Photometric atlas of the solar spectrum from [lambda] 7498 to [lambda] 12016. [Par] L. Delbouille [et] G. Roland.}",
         year = 1963,
       adsurl = {https://ui.adsabs.harvard.edu/abs/1963apss.book.....D}
}

@article{Aboudarham2000,
  doi = {10.25935/9TXJ-F095},
  url = {http://bass2000.obspm.fr/DOI/reference_paper_bass2000.html},
  author = {Aboudarham,  Jean and Renié,  Christian},
  language = {en},
  title = {BASS2000,  database of Solar ground-based observations},
  publisher = {BASS2000/PADC},
  year = {2020},
  copyright = {Creative Commons Attribution Non Commercial Share Alike 4.0 International}
}

@ARTICLE{DeWilde2025,
       author = {{De Wilde}, M. and {Pietrow}, A.~G.~M. and {Druett}, M.~K. and {Pastor Yabar}, A. and {Koza}, J. and {Kontogiannis}, I. and {Andriienko}, O. and {Berlicki}, A. and {Brunvoll}, A.~R. and {de la Cruz Rodr{\'\i}guez}, J. and {Thoen Faber}, J. and {Joshi}, R. and {Kuridze}, D. and {N{\'o}brega-Siverio}, D. and {Rouppe van der Voort}, L.~H.~M. and {Ryb{\'a}k}, J. and {Scullion}, E. and {Silva}, A.~M. and {Vashalomidze}, Z. and {Vicente Ar{\'e}valo}, A. and {Wi{\'s}niewska}, A. and {Yadav}, R. and {Zaqarashvili}, T.~V. and {Zbinden}, J. and {{\O}yre}, E.~S.},
        title = "{Synthesizing Sun-as-a-star flare spectra from high-resolution solar observations}",
      journal = {\aap},
         year = 2025,
        month = aug,
       volume = {700},
          eid = {A275},
        pages = {A275},
          doi = {10.1051/0004-6361/202554870},
archivePrefix = {arXiv},
       eprint = {2507.07967},
 primaryClass = {astro-ph.SR},
       adsurl = {https://ui.adsabs.harvard.edu/abs/2025A&A...700A.275D}
}

@ARTICLE{Fraunhofer1817,
       author = {{Fraunhofer}, Joseph},
        title = "{Bestimmung des Brechungs- und des Farbenzerstreungs-Verm{\"o}gens verschiedener Glasarten, in Bezug auf die Vervollkommnung achromatischer Fernr{\"o}hre}",
      journal = {Annalen der Physik},
         year = 1817,
        month = jan,
       volume = {56},
       number = {7},
        pages = {264-313},
          doi = {10.1002/andp.18170560706},
       adsurl = {https://ui.adsabs.harvard.edu/abs/1817AnP....56..264F}
}

@software{specutils2019,
       author = {{Astropy-Specutils Development Team}},
        title = "{Specutils: Spectroscopic analysis and reduction}",
 howpublished = {Astrophysics Source Code Library, record ascl:1902.012},
         year = 2019,
        month = feb,
          eid = {ascl:1902.012},
archivePrefix = {ascl},
       eprint = {1902.012},
       adsurl = {https://ui.adsabs.harvard.edu/abs/2019ascl.soft02012A}
}

@ARTICLE{Greisen2006,
       author = {{Greisen}, E.~W. and {Calabretta}, M.~R. and {Valdes}, F.~G. and {Allen}, S.~L.},
        title = "{Representations of spectral coordinates in FITS}",
      journal = {\aap},
         year = 2006,
        month = feb,
       volume = {446},
       number = {2},
        pages = {747-771},
          doi = {10.1051/0004-6361:20053818},
archivePrefix = {arXiv},
       eprint = {astro-ph/0507293},
 primaryClass = {astro-ph},
       adsurl = {https://ui.adsabs.harvard.edu/abs/2006A&A...446..747G}
}

@book{babcock1947,
  title={The Solar Spectrum, [lambda] 6600 to [lambda] 13495},
  author={Babcock, H.D. and Moore, C.E.},
  isbn={9780598491794},
  lccn={48002416},
  series={Carnegie Institution Washington, DC: Carnegie Institution of Washington publication},
  url={https://books.google.nl/books?id=cSLVzwEACAAJ},
  year={1947},
  publisher={Carnegie Inst.}
}

@BOOK{Moore1996,
       author = {{Moore}, Charlotte E. and {Minnaert}, M.~G.~J. and {Houtgast}, J.},
        title = "{The solar spectrum 2935 A to 8770 A}",
         year = 1966,
       adsurl = {https://ui.adsabs.harvard.edu/abs/1966sst..book.....M}
}

@misc{Kramida1991,
  doi = {10.18434/T4W30F},
  url = {http://www.nist.gov/pml/data/asd.cfm},
  author = {Kramida,  Alexander and Ralchenko,  Yuri},
  language = {en},
  title = {NIST Atomic Spectra Database,  NIST Standard Reference Database 78},
  publisher = {National Institute of Standards and Technology},
  year = {1999},
  copyright = {License Information for NIST data}
}

\end{document}